\documentclass[fleqn,twoside]{article}
\usepackage{espcrc2,epsfig}

\usepackage{graphicx}
\usepackage[figuresright]{rotating}

\newcommand{\AmS}{{\protect\the\textfont2
  A\kern-.1667em\lower.5ex\hbox{M}\kern-.125emS}}

\title{   \vspace{-3.65cm}
       {\normalsize DESY 01-166}     \\[-0.2cm]      % for preprint
       {\normalsize October 2001}    \\              % for preprint
       \vspace{2.70cm}                               % for 1 preprint #
Separating perturbative and non-perturbative contributions
 to the plaquette%
 \thanks{Talk given by P. Rakow at Lattice 2001, Berlin.}}

\author{ 
 R. Horsley%
\address[NIC]{NIC/DESY Zeuthen,
                    D-15738 Zeuthen, Germany},
  P.~E.~L.~Rakow%
 \address[Rgb]{Institut f\"ur theoretische Physik, 
        Universit\"at Regensburg, D-93040 Regensburg, Germany} 
  and G.~Schierholz\addressmark[NIC]$^,$%
           \address[DESY]{Deutsches Elektronen-Synchrotron DESY,
                          D-22603 Hamburg, Germany}
 }
       
\begin{document}

\begin{abstract}
   We try to separate the perturbative and non-perturbative
 contributions to the plaquette of pure $SU(3)$ gauge theory. 
 To do this we look at the large-$n$ asymptotic behaviour of the 
 perturbation series in order to estimate
 the contribution of the as-yet uncalculated terms in 
 the series.  We find no evidence for the previously
 reported $\Lambda^2$ contribution
  to the gluon condensate. Attempting to determine the conventional
   $\Lambda^4$ condensate gives a value 
  $\sim 0.03(2)$ GeV$^4$, in reasonable
    agreement with sum rule estimates, though with
 very large uncertainties.
\vspace*{-0.3cm}
\end{abstract}

\maketitle

%\normalsize

%----------------------------------------------------------------------------

 \section{INTRODUCTION} 

 There is a long history of trying to extract a non-perturbative
 gluon condensate from lattice calculations of the plaquette~\cite{early}.  
 Thanks to the work of~\cite{DiRenzo} we have a much longer
 perturbation series for the plaquette than for any other 
 lattice quantity. On the other hand, the plaquette is one
 of the most ultra-violet dominated quantities, which means 
 that any non-perturbative contribution is likely to be very
 small compared with the perturbative part. 

  Conventionally one expects that the 
 non-perturbative part of the plaquette is proportional to 
 $a^4$, and related to the gluon condensate introduced in~\cite{SVZ}
 \vspace*{-0.2cm}
  \begin{equation}
  P_{MC} = P_{pert} - a^4 \frac{\pi^2}{6^2} 
 %\left(1\!- \frac{b_1}{b_0} g^2\!+ \cdots \right)
 \left[\frac{-b_0 g^3}{\beta(g)} \right]
 \left\langle \frac{\alpha}{\pi} G G \right\rangle  
 \label{a4_cond}
 \end{equation}
 where $b_0$ is the first coefficient of the $\beta$-function. 
 However one study has reported a non-perturbative
 contribution scaling like $a^2$~\cite{Burgio}. 

   The Monte Carlo quenched plaquette, $P_{MC}$,
 can be measured very accurately. The difficult part in determining
 the condensate is finding the sum of the perturbative
 series $P_{pert} \equiv 1 - \sum_n p_n g^{2 n}$. 
 (We use $g^2$ as our expansion parameter, rather than $\beta^{-1}$.)
 We know the first few terms in this series from conventional
 lattice perturbation theory calculations. Di~Renzo et~al.~have 
 managed to calculate many more terms in the series by a clever
 stochastic method~\cite{DiRenzo}.
 Since this is the only lattice quantity 
 with a long perturbation series, it gives us a unique 
 opportunity to see how the coefficients behave at large $n$. 
  The calculations, which were 
 done on a $24^4$ lattice, reproduced the three known
  terms correctly. 
% \vspace*{-0.5cm}

 \section{EXTRAPOLATING THE SERIES}

    To estimate any non-perturbative contributions to the plaquette,
 we need to know the sum of the perturbation series to one part in
 $10^3$ or better. We quickly see that for interesting couplings
 ($g^2 \approx 1$) 10 loops is not enough, and that we need a
 good estimate of the contribution from higher order terms.
 
   To extrapolate to higher loops, we need to know how the 
 coefficients depend on $n$. A good way of determining this behaviour
 is to look at $r_n$, the ratio of adjacent coefficients. 
 In statistical mechanics, $r_n$ is
 often plotted against $1/n$ rather than $n$. This is because
 this gives a straight line for a series with a
 power-law behaviour,
 \vspace*{-0.2cm}
   \begin{eqnarray}
 \lefteqn{(1 - u x)^q =} \\
 && \! \! 1 - q u x + \cdots
 + \frac{\Gamma(n-q)}{\Gamma(n+1) \Gamma(-q)} (u x)^n + \cdots
 \nonumber
 \end{eqnarray}
 so that
 \clearpage
  \begin{equation}
 %\frac{c_n}{c_{n-1}} = u \left(1 - \frac{1+q}{n}\right)
   c_n /c_{n-1}  = u \; [1 - (1+q)/n ]
 \end{equation}
 where the $c_n$ are the coefficients in the Taylor expansion.

 \begin{figure}[t]
% \vspace*{-0.8cm}
\begin{center}
\epsfig{file = 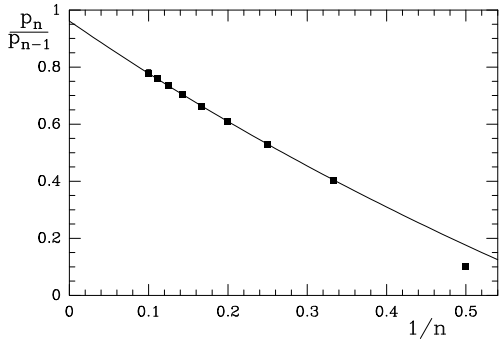, angle = 270, width = 6.5cm}
\end{center}
 \vspace*{-0.7cm}
\caption{\it The ratio of successive coefficients in the perturbation
 series of the plaquette, plotted against $1/n$.
 The solid line is the fit of eq.(\ref{pow_fit}).
 \label{over_n}}
 \vspace*{-0.7cm}
 \end{figure}

 The plot, Fig.~\ref{over_n},
 is almost linear, though there is a little curvature,
 which is taken into account by making a fit of the form
% \begin{eqnarray} 
% \lefteqn{
% \frac{p_n}{p_{n-1}} = u \left(1 - \frac{1+q}{n+s}\right)\;,}
% \label{pow_fit} \\
% &u = 0.961(9), &\! q=0.99(7),\ \ \ \ s =0.44(10) \;.\nonumber
% \end{eqnarray}
 \begin{equation} 
 r_n \equiv \frac{p_n}{p_{n-1}} = u \left(1 - \frac{1+q}{n+s}\right)\;, 
 \label{pow_fit} 
 \end{equation}
 with $u = 0.961(9)$, $q=0.99(7)$, $s =0.44(10)$.
 This series converges for $|g^2| < |u|^{-1}$. 
% The sum of the series gives the hypergeometric function 
% ${}_2 F_1(s-q,1,s+1,u g^2)$. 
 At first sight it might seem surprising that a series with
 a power-law singularity describes the data well. However, it is
 known that the specific heat of lattice gauge theories has a sharp
 peak in the cross-over region between strong and weak coupling  
 (for an example, see~\cite{Lautrup}). If this feature dominates the
 series, a fit like eq.(\ref{pow_fit}) is reasonable. 

\begin{figure}[t]
 \vspace*{0.1cm}
\begin{center}
\epsfig{file = 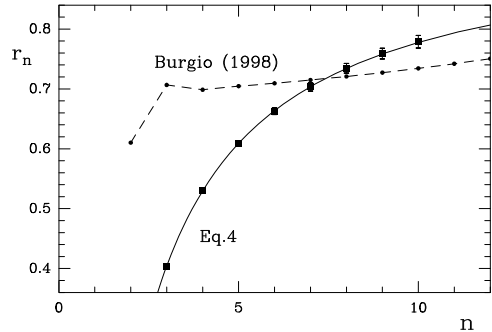, angle = 270, width = 6.35cm}
\end{center}
\vspace*{-0.7cm}
\caption{\it The ratio $r_n$ compared with the renormalon fit
 of~\cite{Burgio} (dashed line)
 and the fit of eq.(\ref{pow_fit}) (solid line). 
 \label{ren_fit_plot} }
 \vspace*{-0.8cm}
 \end{figure}

   To estimate the higher-order terms in the series,~\cite{Burgio} use
 a different extrapolation, 
 based on renormalon ideas about the asymptotic form of
 the series. Their formula for the coefficients is quite complicated,
 for full detail, see~\cite{Burgio}. They assume that there is a 
 scheme in which renormalon behaviour (factorial growth of the 
 coefficients) applies, and that the coupling in 
 this scheme is related to the usual lattice coupling by a transformation
 of the type 
  \begin{equation}
 \frac{1}{g^2_{\rm ren}} =  \frac{1}{g^2} -\frac{r}{6}
 -\frac{r^\prime}{6^2} g^2 \;.
  \end{equation}
 Using the coefficients known at the time (up to $n=8$) Burgio
 et~al.~made a fit for 
 $r$ and ${r^\prime}$ which leads to the dashed line shown in 
 Fig.~\ref{ren_fit_plot}. Although this renormalon
 fit has about the right value near $n \approx 8$, it does not
 reproduce the $n$-dependence as well as the fit eq.(\ref{pow_fit}).  

\begin{figure}[htp]
 \vspace*{-0.8cm}
\begin{center}
\epsfig{file = 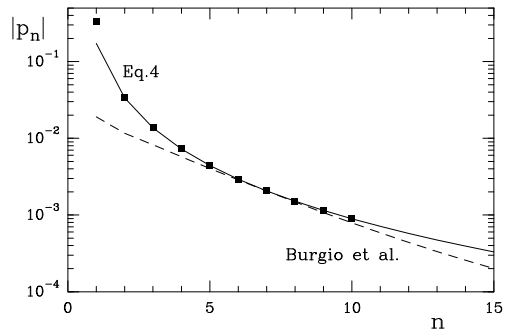, angle = 270, width = 6.5cm}
\end{center}
 \vspace*{-0.8cm}
\caption{\it The coefficients in the plaquette series, compared
 with the fit of~\cite{Burgio} (dashed line) and of eq.(\ref{pow_fit})
 (solid line). \label{p_plot}} 
 \vspace*{-0.9cm}
 \end{figure}

 The resulting fit for the $p_n$ is shown in Fig.{\ref{p_plot}},
 and compared with the renormalon-inspired fit. 
 Eq.(\ref{pow_fit}) describes the coefficients from 2 loops
 onward very well, while the renormalon fit shows much less
 curvature than the data. 
 The new extrapolation predicts larger coefficients for the unknown
 terms in the perturbation series. That means that there will be
 a smaller condensate left after subtracting $P_{pert}$.

 \section{THE GLUON CONDENSATE}

 Using the 8 coefficients which they had then, and estimating the
 remainder of the perturbative series from their renormalon
 fit,~\cite{Burgio} produced values for the non-perturbative 
 part of the plaquette for $\beta$ between 6.0 and 7.0. 
 The surprising result was that $\Delta P$, the difference between the 
 perturbative plaquette and the Monte Carlo result, was not proportional
 to $a^4 \Lambda^4$ as conventionally expected, 
 but to $a^2 \Lambda^2$. 

 \begin{figure}[b]
 \vspace*{-0.7cm}
\begin{center}
\epsfig{file = 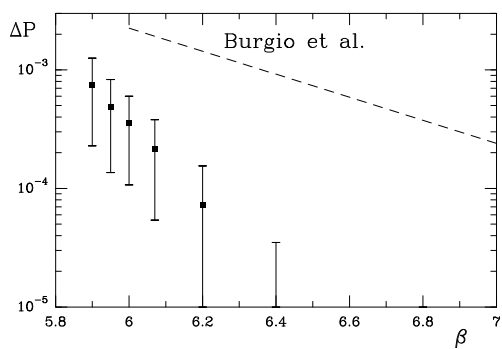, angle = 270, width = 6.5cm}
\end{center}
\vspace*{-0.8cm}
\caption{\it $\Delta P$, the difference between the Monte Carlo
 and perturbative plaquette, according to this work (points),
 and according to~\cite{Burgio} (dashed line).
 \label{log_condens}}
 \end{figure}

 In Fig.~{\ref{log_condens}}
 we compare $\Delta P$
 %, the difference between the Monte Carlo
 %plaquette and the perturbative result, 
  using our estimate of the perturbative
 plaquette, and using the estimate from~\cite{Burgio}.  We see that we have 
 a far smaller condensate left over, and that there is 
 no longer much support for the unconventional $a^2$ slope.  

 \begin{figure}[t]
\begin{center}
\epsfig{file = 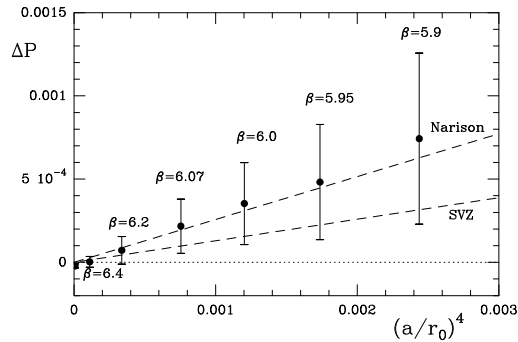, angle = 270, width = 6.7cm}
\end{center}
\vspace*{-0.8cm}
\caption{\it  $\Delta P$, compared with phenomenological values
 of the gluon condensate
 from~\cite{SVZ,Narison}.
 \label{phenom} }
 \vspace*{-0.6cm}
 \end{figure}

 In Fig.~{\ref{phenom}} we see that $\Delta P$  
  % the remainder after subtracting off the perturbative contribution, 
 scales roughly as $a^4$, as expected from eq.(\ref{a4_cond}).
 However the errors 
 (mainly from the uncertainty in summing the perturbation series) are 
 very large.
   The dashed lines show the values of the gluon condensate
 expected according to various sum-rule
 estimates~\cite{SVZ,Narison}. 
 % $r_0$ is approximately 0.5$fm$. 
 Using the force scale $r_0$ to set our length 
 scale gives 
% $ \left\langle \frac{\alpha}{\pi} G G \right\rangle
% \sim 0.03(2)$ GeV$^4$,
\vspace*{-0.10cm}
 \begin{equation}
  \left\langle \frac{\alpha}{\pi} G G \right\rangle
 \sim 0.03(2)\; {\rm GeV}^4,
 \end{equation}
\vspace*{-0.10cm}
  which is similar 
 to phenomenological estimates such as
 $\sim 0.012$ GeV$^4$~\cite{SVZ}
 or 0.024(8) GeV$^4$~\cite{Narison}.

 \section{CONCLUSIONS}  

  We can see from Figs~\ref{over_n}-\ref{p_plot} that
 the power law fit, eq.(\ref{pow_fit}), 
 describes the known perturbative coefficients 
 better than the renormalon fit~\cite{Burgio}. 

  If we use eq.(\ref{pow_fit}) to estimate the sum of the 
 perturbative series, we find a number very close to the full
 plaquette as measured in Monte Carlo calculations. So 
 any non-perturbative contribution to the plaquette is very small. 

  The non-perturbative contribution to the 
 plaquette does not look proportional to $a^2 \Lambda^2$, 
 but rather appears to be proportional to $a^4 \Lambda^4$,
 with about the same order of magnitude as the phenomenological
 gluon condensate. 

  Asymptotically, the renormalon behaviour could still be right. 
 The renormalon asymptote predicts $r_n \approx n b_0/2$, which
 does not catch up with eq.(\ref{pow_fit}) until $n \approx 25$. 
 \vspace*{0.3cm}

 {\bf Acknowledgements:} This work was supported
 by the DFG and the BMBF.

\end{document}